# Structuring Stress for Active Materials Control

Rui Zhang[1†], Steven A. Redford[2,3†], Paul V. Ruijgrok[4], Nitin Kumar[3], Ali Mozaffari[1], Sasha Zemsky[5], Aaron R. Dinner[3,6], Vincenzo Vitelli[3,7], Zev Bryant[4,8], Margaret L. Gardel[1,3,7*], and Juan J. de Pablo[1,9*]

[1] Pritzker School of Molecular Engineering, The University of Chicago, Chicago, Illinois 60637, USA.
[2] The graduate program in Biophysical Sciences, The University of Chicago, Chicago, Illinois 60637, USA.
[3] James Franck Institute, The University of Chicago, Chicago, Illinois 60637, USA.
[4] Department of Bioengineering, Stanford University, Stanford, California 94305, USA.
[5] Program in Biophysics, Stanford University, Stanford, California 94305, USA.
[6] Department of Chemistry, The University of Chicago, Chicago, Illinois 60637, USA.
[7] Department of Physics, The University of Chicago, Chicago, Illinois 60637, USA.
[8] Department of Structural Biology, Stanford University Medical Center, Stanford, California 94305, USA.
[9] Center for Molecular Engineering, Argonne National Laboratory, Lemont, Illinois 60439, USA.

[†] equal contribution.

[*] correspondence to gardel@uchicago.edu or depablo@uchicago.edu

**Abstract:** Active materials are capable of converting free energy into mechanical work to produce autonomous motion, and exhibit striking collective dynamics that biology relies on for essential functions. Controlling those dynamics and transport in synthetic systems has been particularly challenging. Here, we introduce the concept of spatially structured activity as a means to control and manipulate transport in active nematic liquid crystals consisting of actin filaments and light-sensitive myosin motors. Simulations and experiments are used to demonstrate that topological defects can be generated at will, and then constrained to move along specified trajectories, by inducing local stresses in an otherwise passive material. These results provide a foundation for design of autonomous and reconfigurable microfluidic systems where transport is controlled by modulating activity with light.

Soft materials in which individual components convert ambient free energy into mechanical work are commonly referred to as active matter (*1*, *2*). These systems are compelling in that their relatively simple rules of propulsion and inter-particle interactions can give rise to intriguing collective behaviors and pattern formation across length scales (*3*, *4*). Active components underpin coherence in a wide range of natural processes. They play a critical role in cellular migration, flocking, and long-range flows in dense bacterial suspensions (*5–9*). These behaviors are not just interesting in and of themselves, but hold promise as the basis for the design of novel, functional materials (*10*, *11*). The central challenge of engineering functionality in active materials is that active flows are often turbulent (*12–16*). Efforts to control these flows thus far have utilized physical boundaries to constrain the material, and rely on spontaneous symmetry breaking to yield steady states and coherent dynamics over a large scale (*17–21*). While these works have demonstrated that a degree of control in active systems is indeed possible, the dependence on physical barriers and spontaneous symmetry breaking limits the amount of control that can be exerted. We seek a different control parameter with which we may direct the flow and dynamics of an active material without the malice of forethought. Ideally, this more flexible control parameter could direct the material asymmetrically and thus allow for the programming of more complex behaviors in active systems. In this letter, we introduce spatially dependent activity as this flexible control parameter, and demonstrate both in experiment and simulations how it can be leveraged to direct defect dynamics and control long-range flows in an active nematic liquid crystal.

Nematic liquid crystals (nematics) are a phase of matter in which extended components—mesogens—align along their long axis to form a material with long-range orientational order, but which can flow like a liquid (*22*). Structural disorder in these systems is stored in distinct regions of discontinuity termed topological defects (*22*). In two dimensions, topological defects carry a "charge" of either + or – ½, defined by the winding number about the defect core (Fig. 1A). When extensile stress is introduced along the orientation of the mesogens, the asymmetric +½ defects are propelled along their axis of symmetry (*23*). The interplay between the active and elastic stress leads to a steady state nucleation, motion and annihilation of defects, resulting in a state known as "active turbulence" (*13*, *24*, *25*). One useful class of active liquid crystals is those formed by cytoskeletal polymers (*12*, *26*). Activity is readily introduced to these systems via the addition of molecular motor proteins that slide adjacent filaments past each other, thereby generating extensile stresses along the nematic orientation (*27*). In this work we construct a nematic liquid crystal by crowding short (~1 μm in length) actin filaments (F-actin) onto an oil-water interface (*28*) (Fig. 1B). Because of the fluorescent dye and the polarized laser used in these experiments, filaments which are vertical in the experimental frame appear brighter than those which are horizontal. Thus, pixel intensity tells us about the local orientation of the nematic field (*29*, *30*) (Fig. 1A).

While previous realizations of cytoskeletal liquid crystals have harnessed the power of naturally occurring motor proteins (*12*, *19*, *26*), in this work we produce spatially structured activity by exploiting motor proteins engineered with light-dependent gliding velocities (*31*). The light-activated gear-shifting myosin motors used here are constructed from myosin XI catalytic heads and a lever arm containing the light sensitive LOV2 domain. The stimulated unfolding of this LOV2 domain changes the geometry and effective length of the lever arm, conferring optical modulation of motor gliding velocity on the biopolymer F-actin (*32*). This geometry allows for advantageous kinetics when compared to previous realizations of light-stimulated activity, which relied on dimerization or inhibitor deactivation to induce rearrangement in cytoskeletal assemblies (*33*, *34*). To generate local stress on antiparallel F-actin, engineered oligomerization domains are utilized to create motor tetramers (*32*, *35*) (Fig. 1C). These tetramers, when added to actin liquid crystals, produce higher defect densities and a greater average nematic speed upon stimulation (Fig. S1, Movie S1). To target this increase in activity to just one region of the liquid crystal a micro-mirror array is used to selectively target the stimulation wavelength of 470 nm to one portion of the sample, while confocal fluorescence imaging is used to visualize fluorescently-tagged actin.

We selectively illuminate a large region (~2000 μm$^2$) (Fig. 1D, red box) in a liquid crystal containing gear-shifting motors. Upon stimulation, the density of topological defects remains low outside of the stimulated region while the activity within leads to defect proliferation (Fig. 1D, yellow circles). Interestingly, this transition from low to high defect density is relatively sharp, occurring within several micrometers of the boundary (Fig. 1E). Moreover, the nematic spontaneously flows within the illuminated region, with an instantaneous velocity that is 3-fold larger than that outside the bounds (Fig. 1F). That defect density and nematic velocity sharply decrease as one leaves the stimulated region implies a sort of confinement of activity within the stimulated region. This confinement can be visualized by tracking the location of +½ defects over time. These trajectories are largely contained within the illuminated region over 400 s, only rarely crossing over from one region to another (Fig. 1G, Movie S2). This observation holds promise for engineering applications, particularly for the control of individual defect dynamics. Consider the example in Fig. 1H; as the defect approaches the border it is deflected. That is, it undergoes a sharp reorientation such that it never crosses the boundary. This implies that a judicious choice of border geometry could allow for the design of motile defect trajectories. This key observation motivated us to explore the extent to which patterned activity could be harnessed to control the proliferation and deflection of defects in active nematics. With this, we envision the capability to arbitrarily pattern active flows and manipulate transport.

The above experimental observations demonstrate how the additional force induced by a gradient in active stress can be used to spatially confine topological defects. To further understand and utilize such additional force to precisely control defects, we turn to comprehensive hydrodynamic simulations of active liquid crystals. Our model is based on a Q-tensor representation of the nematic liquid crystal that incorporates hydrodynamic interactions (*36, 37*). Activity α is introduced as a local force dipole such that the active stress in an incompressible active liquid crystal (*38*) is $\mathbf{\Pi} = -\alpha \mathbf{Q}$. Previously, we and others have considered α to be constant (*23, 26, 27, 37, 39*). We now consider α as a spatial variable, which gives rise to a new stress term due to the gradient of α. Here, a hybrid lattice Boltzmann approach is used to solve the governing equations (SI Text). This method has been shown to be successful in capturing active nematic behaviors over a range of activities (*26, 39*), including the high activity "active turbulent" regime in which topological defects are continuously generated, propelled and annihilated to generate chaotic-like flows (*13, 24, 25*). Here, we use this technique to explore how spatial variation in activity can be used as a tool to control active matter.

We first consider a nematic comprised of two regions of differing activity, $\alpha_1$ and $\alpha_2$, with a flat interface at *x*=0. For $x < 0$ the nematic has a uniform activity of magnitude $\alpha_1$ and for $x > 0$, the activity is $\alpha_2$. Fig. 2A and 2B are snapshots of the dynamic steady state configurations of the nematic order (lines) and instantaneous velocity, respectively, for simulations with $\alpha_1 = 0.0001$ and $\alpha_2 = 0.005$. All simulation data are shown in lattice units where the unit length is chosen to be the mesogen length (SI). For $x \ll x_1$ and $x \gg x_2$, the defect density and velocity approach those expected for an active nematic with uniform activity equivalent to $\alpha_1$ and $\alpha_2$, respectively (Fig. 2C). As in experiment, the trajectories of +1/2 defects created in the $x > 0$ region rarely cross into the low activity region $x < 0$ (Fig. 2D).

To consider the transition between these two regions, we plot the spatial profile of the both +½ and -½ defects across the interface (Fig. 2C). In simulations where $\alpha_1 = 0.0001$ and $\alpha_2 = 0.005$ the density profile for -½ topological defects exhibits a pronounced peak near the interface (Fig. S2). In contrast, the distribution function for +½ defects is flatter and extends into the less active side. The accumulation of defects at the interface gives rise to a topological-charge dipole moment, similar to a recent theoretical calculation for a dry active nematic (*40*), and not unlike that encountered at the interface between charged species of different dielectric permittivity (*41*). As defects try to cross from a high ($\alpha_2$) to a low active ($\alpha_1$) region, they lose mobility and experience an elastic attraction from the opposite-charge defects in the active region. This prevents them from straying deeper into the low-activity side. To quantify the sharpness of these defect density distributions, we identify the transition region, $w_p = x_2 - x_1$, the ends of which are those points where defect density deviates from that expected for a nematic with uniform activity (Fig. 2D). This width, $w_p$, is effectively a measure of the confinement induced by the difference in activity at the interface.

To determine how changes in relative activity impact confinement, we explore how the width of the transition region varies as a function of $\alpha_1$, for simulations with fixed $\alpha_2 = 0.01$. We find that $w_p$ is of the same order of magnitude, and approximately 100-fold that of the nematic coherence length, for all relative activities less than 0.5 (Fig. 2E). As $\alpha_1$ approaches $\alpha_2$, the interfacial width $w_p$ increases (Fig. 2E, S2). Likewise, for a given set of activities, the interfacial width is also quite sensitive to the friction, increasing as the friction is decreased (Fig. S4). Thus, both friction and relative activity can be tuned to construct a sharp interface for defect confinement.

In simulations, we also find that the emergence of defect confinement is accompanied by a preferential mesogen orientation perpendicular to the boundary, creating a so called "anchoring effect" driven by activity gradients. As seen in Fig. 2A, the directors on the low-activity side adopt a normal orientation to the interface. To characterize the anchoring near and at the interface, we define an order parameter, $P_2(\mathbf{n} \cdot \mathbf{v}) = \langle 3(\mathbf{n} \cdot \mathbf{v})^2 - 1 \rangle/2$, where $\mathbf{n}$ is the director field, $\mathbf{v}$ is the interfacial normal at both $x = 0$ and $x_1$, and $\langle \rangle$ denotes an ensemble average. At $x = 0$, no anchoring is observed for any relative activity levels (Fig. 2F, blue squares). However, at $x = x_1$, normal anchoring becomes prominent for relative activities less than 0.1 when the interface is prominent (Fig. 2F, red triangles). Thus, a sharp gradient in activity simultaneously constrains defects to the region of higher activity and anchors the director field in the low activity region in the direction normal to the interface. Together, these results further suggest that structured activity is a means to control nematics regionally, at scales much larger than the defect spacing, potentially providing more flexibility than that previously demonstrated with physical barriers (*19, 20*).

We next use simulations to explore the minimum length scale at which structured activity can be used to manipulate liquid crystals. In particular, the extent to which spatially structured activity can be utilized to create and manipulate defects. In nematics with homogeneous activity, defect creation arises from instabilities in bending undulations (*14, 42*). The level of active stress sets the undulation wavelength and, therefore, sets a length scale required for defect nucleation (*23*). We first consider the effects of adding activity ($\alpha = 0.03$) within a rectangular region, with dimensions slightly larger than this natural length scale, in an initially uniform nematic (Fig. 3A, inset). Because of symmetry, two pairs of defects are created simultaneously (Fig. 3A, Movie S3). While creating defects in this manner is promising, we desire asymmetric control such that we can create single pairs of defects. To achieve asymmetry, we consider a triangular region with base *b* and height *h* of similar dimensions to the rectangle. Here, activity-induced bending instabilities incline towards the triangle tip and lead to the formation of a single pair of ±1/2 defects. To determine how defect pair creation depends on triangle size and activity level, we perform simulations over a range of activities and pattern size *b*, for a given aspect ratio *h/b*=3. For a given size *b*, we map out the threshold activity required to generate a defect pair (Fig. 3C, Movie S4). When *b*>50, the pattern becomes sufficiently large that it surpasses the bending undulation wavelength in a uniform nematic (Fig. 3C, top axis). Here the threshold level of activity, $\alpha_0$, required to generate a defect pair is similar to that found in a nematic with homogeneous activity and, as may be expected, more than one defect pair can be created. For $b < 50$, the activity level required to generate a defect pair increases as the required length scale decreases (Fig. 3C, red triangles). However, we find the stress needed is less than would be required for defect generation in the absence of structured activity. Indeed, the activity gradient creates an additional stress that contributes to defect nucleation. By changing the geometry of the pattern, the threshold activity of defect generation can be varied (Fig. S3). Thus, a judicious choice of activity and geometry allows for control of the nematic field at the scale of individual topological defects.

Having demonstrated the potential for control over defect creation, we next use simulations to consider the extent to which local activity gradients can control the movement of pre-existing defects. First, we consider a passive nematic in which a +½ defect is oriented towards a -½ defect and separated by a distance $d=250$, as shown in Fig. 4A. With this geometry, a low amount of uniform activity ($\alpha = 0.2\alpha_0$) induces the horizontal motion of the +½ defect due to the asymmetric distribution of active stress (23). Eventually, this leads to annihilation of the defect pair (Movie S5). It is therefore of interest to explore how activity gradients could drive motion that deviates from this behavior. Using the same initial conditions, we selectively activate a rectangular region of dimension 290×80 around the +½ defect and consider the effect of rotating the rectangle by an angle $\phi$. In Fig. 4A, we show a time sequence of the simulations for $\phi = 45°$ and show that the +½ defect re-orients to follow the long axis of the rectangle and deflects its trajectory. This is consistent with the defect deflection observed experimentally (Fig. 1G). Next, we explore how varying angle $\phi$ impacts defect trajectories; we find that defects are faithfully guided up to a threshold angle of 60° (Fig. 4B). Above this, defects are no longer reoriented by the patterned activity (Fig. 4C, purple triangles). We then consider how this threshold angle depends on active stress by systematically varying activity. Because an activity value greater than $\alpha_0$ will result in defect creation and not simply redirection, we consider only activities that are less than $\alpha_0$. When the activity is increased from $0.2\alpha_0$ to $0.3\alpha_0$ and $0.5\alpha_0$, the threshold angle decreases to 45° and 30°, respectively. This can be understood both by the increased defect speed at higher activities and by the effect of activity on local bend distortion that limits the reorientation of the +½ defect. Thus, as the structured activity approaches $\alpha_0$, the ability to manipulate individual +½ defects becomes limited. Together with Fig. 3, these data demonstrate how the shape and magnitude of structured activity can be exploited for individual defect generation and manipulation.

To experimentally test for control over individual defects, we construct quarter-annulus regions in which we stimulate the local activity. We start with a +½ defect at the top left (Fig. 5A). Using an activity well below $\alpha_0$ (Fig. S1E-F), we find that an isolated defect is re-oriented as it follows the pattern. This results in the defect traveling to the other side of the pattern, and pointing 90° from its initial alignment (Fig. 5A, Movie S6). This behavior can be seen in a number of independent samples (Fig. 5B). The rotation angle probability density function (PDF) of directed defects on the pattern shows a pronounced peak near $\theta = 25°$ (Fig. 5C) which indicates that for the five independent defects considered here, the annulus imposes a more or less constant turn angle (43). This is in sharp contrast to a similar histogram constructed from the trajectories obtained from Fig. 1F, which show a relatively uniform distribution (Fig. 5D). That the trajectories from the annulus pattern produce a sharply peaked angle change distribution while a large stimulated region begets a relatively flat PDF evidences further that defects are constrained and directed by the pattern.

To envision how these principles can be applied in microfluidic systems, we demonstrate control of defect pathways by designing specific activity patterns in simulations. Specifically, we consider an "H" channel with two T-junctions containing a defect-free nematic adopting normal anchoring to the surfaces (Fig. 6A-C). We first use a triangular pattern to generate a ±½ defect pair (Fig. 6A). Note that if activity is added uniformly through the nematic, the defect will move into the top right channel (Fig. S5). To try and exert influence on this fate we consider two different activity patterns: a narrow stripe that is horizontal (Fig. 6B, Movie S7) or a "z" pattern (Fig. 6C, Movie S7). The +½ defect is directed along either activity pattern, ultimately being directed into either the top or bottom channel. As these scenarios evolve, the nematic organization is deformed and we can consider how the total elastic energy of the system changes as the defect moves (Fig. 6D). We find the total elastic energy of the final state in the case of stripe pattern is lower than that for the z-pattern. Thus, in simulation, structured activity is a means by which to control defect fate in a microfluidic system despite distinct fates being uphill with respect to energy.

Spatially structured activity presents a promising direction for engineering structure and transport in active matter at many length scales. For one, defects, and the flows they generate, can be confined on a scale larger than the average defect spacing. This results in steady-state defect density distributions and promises the types of confined flows seen throughout this work. However, the main strength of this approach is not its ability to merely control bulk flow, but its specificity at a smaller scale. The number and distribution of defects are key state variables of any nematic system. The ability to specifically nucleate a single defect pair and similarly to be able to manipulate the positions of preexisting defects are steps along the road to controlling these variables. One could imagine composing these two operations spatially to arbitrarily control the entire nematic director. On a wider scale, the flexibility of spatially structured activity is what excites the most when thinking of its applications in active systems in general. The ability to exert control in both space and time and across length scales opens the door for programming complex behaviors into active systems. One can imagine temporal and spatial control of activity working together to achieve complex transport tasks or induce novel non-equilibrium stead states. Much work has yet to be done in order to exert such truly multi-scale control, but we hope that the results presented here may serve as groundwork for future endeavors.

**Acknowledgments:** R.Z., S.R., N.K. and A.M. thank helpful discussions with M. Cristina Marchetti, Suraj Shankar, and Sriram Ramaswamy. R.Z. and A.M. are grateful for the support of the University of Chicago Research Computing Center for assistance with the calculations carried out in this work. This work is primarily supported by the University of Chicago Materials Research Science and Engineering Center, which is funded by the National Science Foundation (NSF) under

Award DMR-1420709. J.J.d.P. acknowledges support from NSF Grant DMR-1710318. N.K. acknowledges the Yen Fellowship of the Institute for Biophysical Dynamics, The University of Chicago. Z.B. acknowledges support from NIH R01 GM114627 and a W.M. Keck Foundation grant to Z.B. and M. Prakash. S.R. acknowledges support from the University of Chicago Materials Research Science and Engineering Center (MRSEC). R.Z., S.A.R., N.K., M.L.G., and J.J.dP conceived the research. R.Z. performed simulations. S.A.R. performed experiments. P.V.R., S.Z., and Z.B. provided reagents. R.Z. and S.A.R. performed data analysis. N.K., A.M., V.V., and A.R.D. contributed to analysis and interpretation. M.L.G. and J.J.dP supervised the research. R.Z., S.A.R., M.L.G, and J.J.dP wrote the manuscript. Everyone contributed to the discussion and manuscript revision.

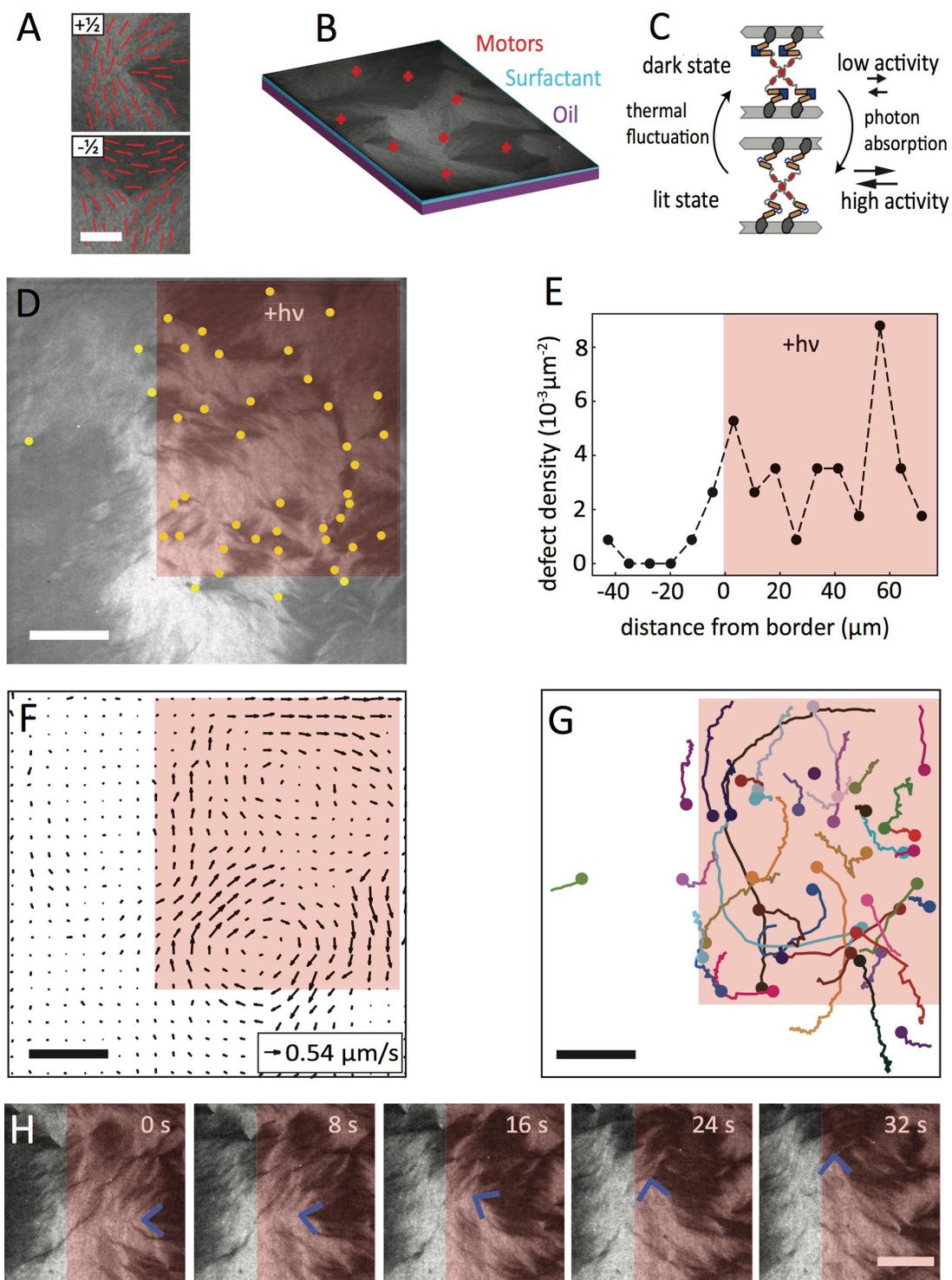

**Fig. 1. Patterning activity in an actin liquid crystal leads to spatially confined flows and topological defects**. A) Polarization image of actin filaments forming a +½ (top) and -½ defect (bottom). Brighter (darker) pixels in the image are regions in which filaments are vertical

(horizontal). Red lines indicate the local average orientation of filaments (director). B) A schematic of the experimental setup. Actin is crowded onto an oil-water interface by methylcellulose (not pictured) where engineered myosin motors generate active stress. C) A schematic illustrating the gear-shifting motors. Upon illumination with 470 nm light, a change in the motor lever arm length leads to higher gliding velocity. D) An experimental polarization microscopy snapshot of fluorescently tagged actin driven by MyLOVChar4~1R~TET. Conditions detailed in Supp. Table 1, "Figure 1". Gear-shifting motors were stimulated only in within the red box. Topological defects as described above are indicated by yellow dots. Scale bar 20 μm. E) Velocity field corresponding to the frame in (D). F) A line-scan of defect density averaged across the y direction as a function of distance from the border of the activated region in (D). G) +1/2 defect trajectories for the first 400 s of stimulation for the experiment excerpted in (D). H) An example of a trajectory in which a defect "deflects" off of the boundary of the stimulated region.

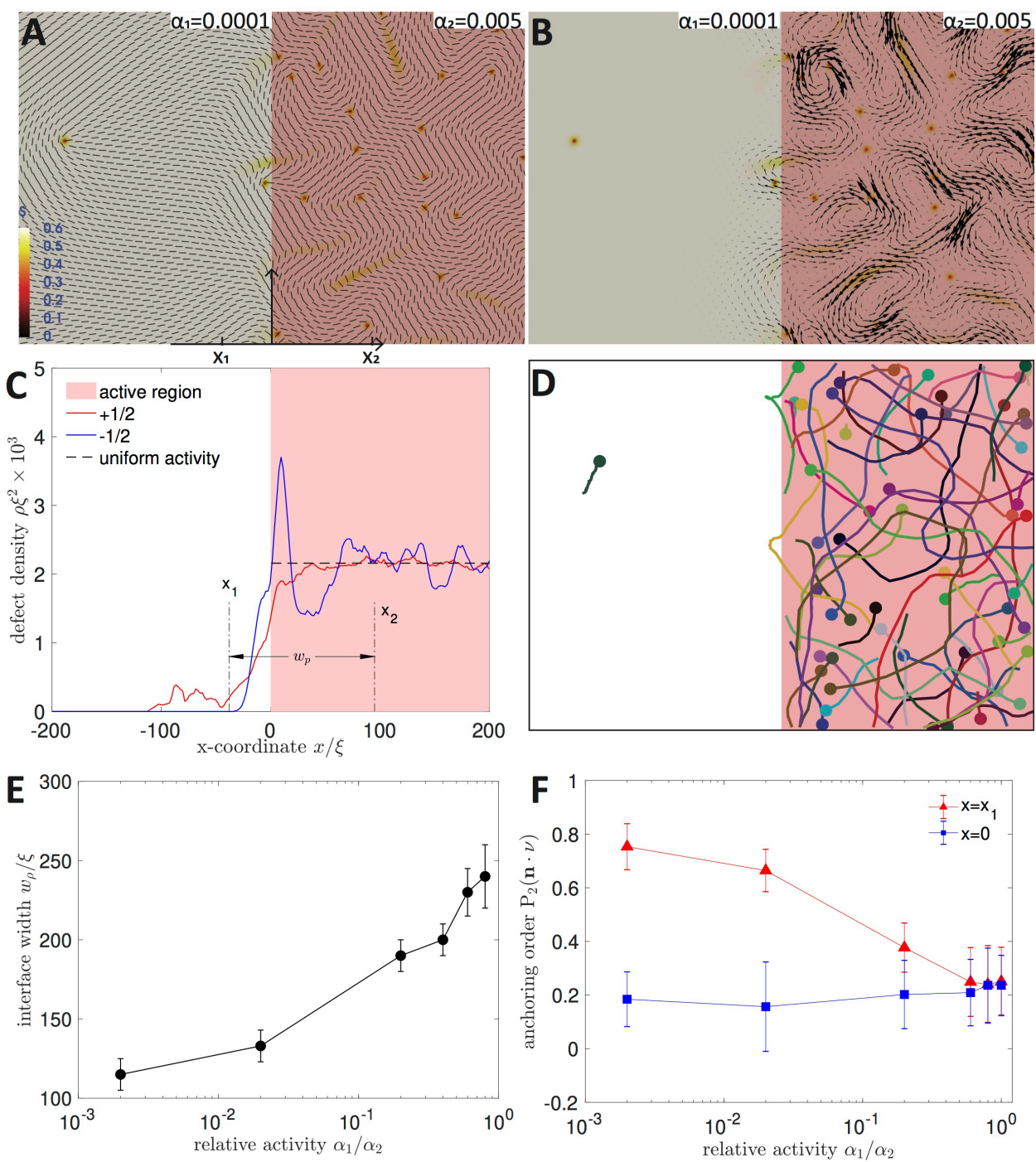

**Fig. 2. Simulations of defect behavior in a patterned active nematic.** A snapshot of the director field (A) for a flat active-less-active boundary located at x=0, its corresponding velocity field (B), defect density profile (C), and defect trajectories (D). Active region is colored with light red in (A-C). The background in (A-B) is colored with nematic order parameter S, with dark red indicating defect locations. (E) Defect density interfacial width $w_p$ as function of relative activity $\alpha_1/\alpha_2$. (F)

scalar order parameter $P_2(\mathbf{n}\cdot\mathbf{v})$ characterizing anchoring effect at different locations with respect to the boundary of activity pattern.

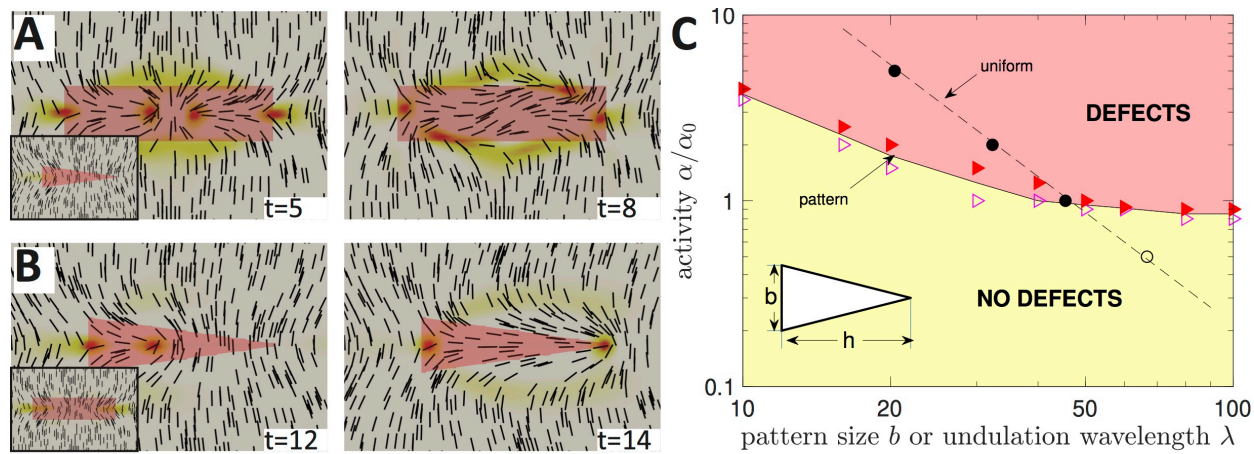

**Fig. 3. Simulations of defect pair creation using activity pattern.** Sequential images of an initially uniform active nematic with a rectangular (A) and a triangular (B) pattern at activity level $\alpha = 3\alpha_0$. Initial configurations are shown in insets of (A) and (B). The triangle (rectangle) has a base (height) b=14 and height (width) h=50. (C) Threshold activity for different pattern sizes at fixed aspect ratio h/b=3. Spontaneous undulation wavelengths at given activity are marked with open (no defect generation) and filled (defect generated) black symbols.

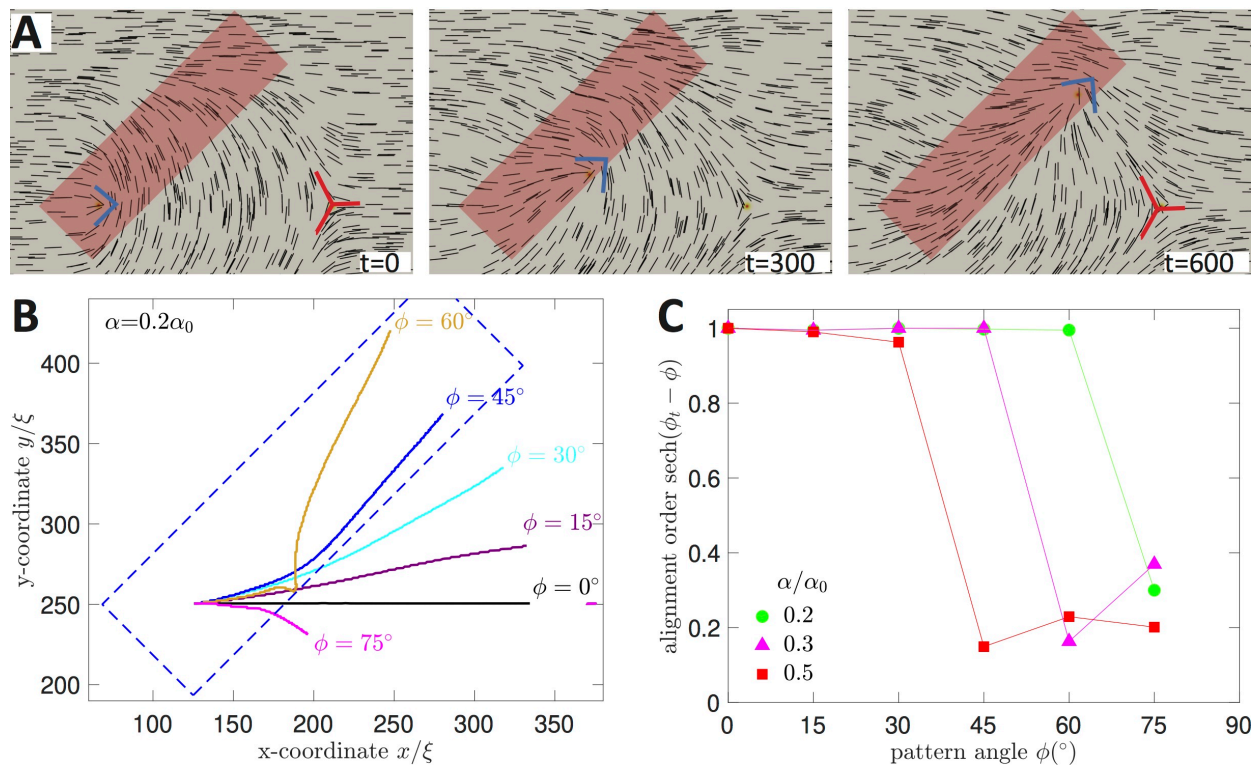

**Fig. 4**. **Simulations of defect deflection by a rectangular activity pattern**. (A) sequential images showing defect deflection at activity $\alpha = 0.2\alpha_0$ and box tilt angle $\phi = 45°$. Rectangle size is [290,80] and initial defect separation is 250. Defects are marked to aid eyes. (B) defect trajectories for different tilt angles at $\alpha = 0.2\alpha_0$. Activity pattern is shown as dashed boxes for $\phi = 45°$. (E) aligning order parameter sech $(\phi_t - \phi)$ as function of imposed angle $\phi$ for various activities with $\phi_t$ the angle of the asymptotic trajectory with respect to +x.

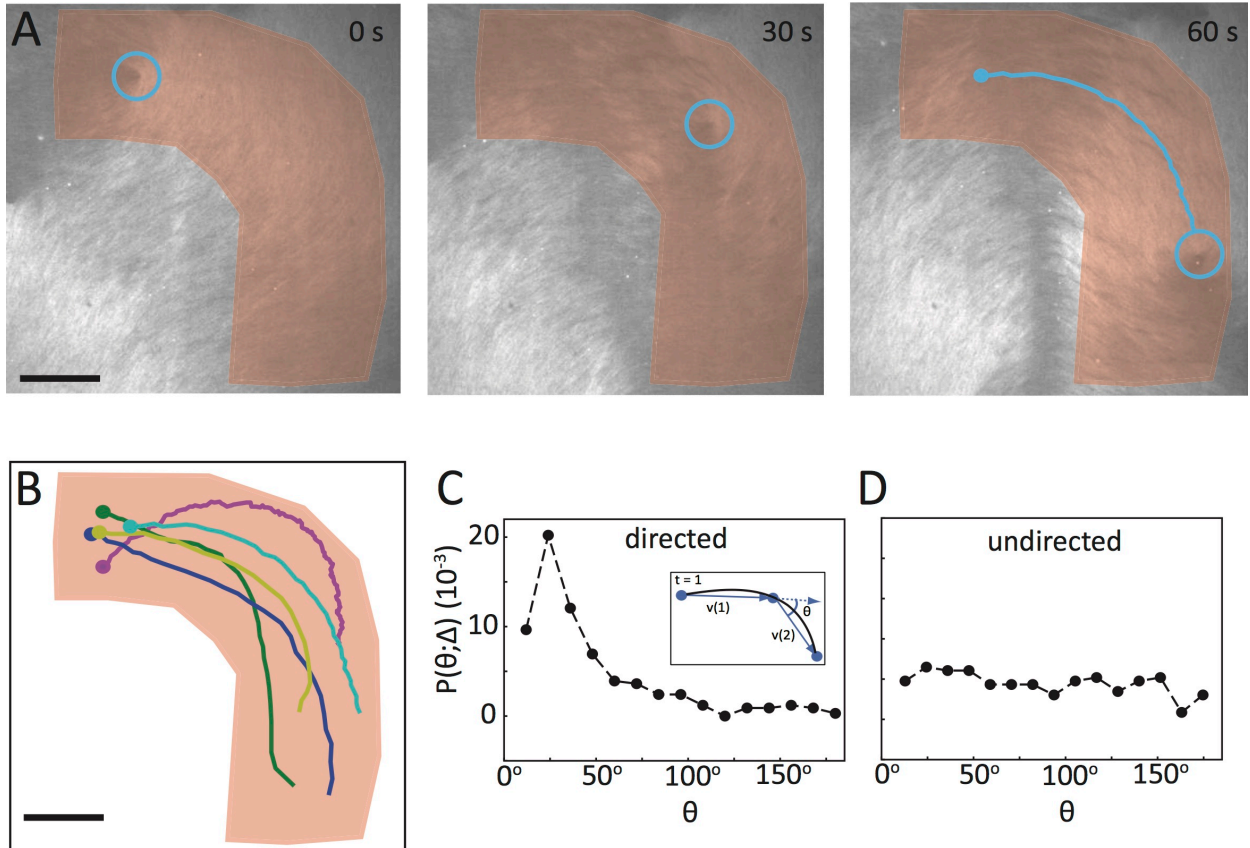

**Fig. 5. Targeted activation can be used to direct defect trajectories in experiment**. A) Series of experimental images showing a plus one-half defect (blue chevron) moving within the active pattern (red outline). The trajectory of the defect is shown as a tail on the defect in the last frame. B) The trajectories for five independent samples from two different days (scatterplot) in relation to the activated region. C) Probability density function (PDF) for a given change in vector angle, $\theta$, for the trajectories in (B). Inset is a schematic of the method for quantifying angle change. Briefly, trajectories are sampled every $\Delta$ steps. The angle between two adjacent vectors is the angle of the dot product of those vectors. The time lag is $\Delta=4$. D) PDF of change in vector angle for the trajectories in Fig. 1F. The time lag is $\Delta=5$ (to account for different frame rate). Scale bars are 20 µm.

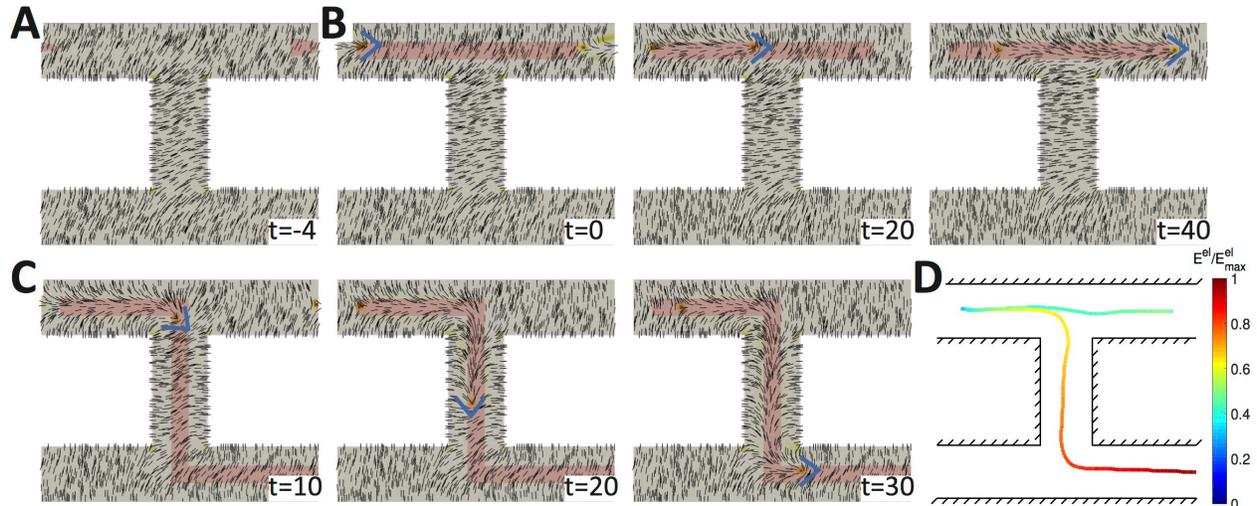

**Fig. 6**. **Simulations of defect pathway control in a channel system**. (A) A triangular pattern with activity $\alpha = 1.8\alpha_0$ is used to create a $\pm 1/2$ defect pair on an otherwise defect-free channel with normal anchoring condition. (B) sequential images showing a $+1/2$ defect moves to the top right channel under a stripe pattern at $\alpha = 0.3\alpha_0$. (C) sequential images showing a $+1/2$ defect moves to the bottom right channel under a "z" pattern at $\alpha = 0.4\alpha_0$. $+1/2$ defect is marked to aid eyes. (D) Defect trajectories colored by the system's elastic energy for the two types of pattern.